\documentclass[aps,pra,preprint,showpacs,preprintnumbers,amsmath,amssymb]{revtex4}
\usepackage{amsmath,mathrsfs,amsbsy,color,graphicx,bm,amsthm,amsfonts}
\usepackage{units}
\usepackage{bbm}
\usepackage{times}
\usepackage{dcolumn}
\usepackage{mathrsfs}
\usepackage{amsmath,amssymb,epsfig}
\definecolor{RED}{rgb}{1,0,0}
\newcommand{\rev}[1]{\textcolor{red}{\textbf{#1}}}
\begin{document}

\title{Dilaton gravity can enhance quantum coherence and reduce entanglement}
\author{Zejun Wang$^1$\footnote{Email: 12022832@czc.edu.cn}, Zhihong Liu$^1$, Yu-Xuan Wang$^2$, Xiao-Li Huang$^2$\footnote{Email: huangxiaoli1982@foxmail.com (corresponding author)}}
\affiliation{$^1$  Department of Physics, Changzhi University, Changzhi, 046011, China \\
$^2$  Department of Physics, Liaoning Normal University, Dalian 116029, China
}


\begin{abstract}

We investigate the influence of the Garfinkle-Horowitz-Strominger (GHS) dilaton black hole on different quantum resources of Dirac fields beyond the single-mode approximation. By employing the negativity to characterize quantum entanglement and the $l_1$-norm and the relative entropy of coherence to characterize quantum coherence, we demonstrate that these resources exhibit remarkably different responses to the gravitational field. Specifically, increasing the dilaton parameter continuously suppresses quantum entanglement, leaving only a finite residual amount in the strong-gravity regime, whereas quantum coherence is enhanced, indicating that the dilaton-induced spacetime affects nonlocal quantum correlations and local quantum superposition in fundamentally different ways. Furthermore, we show that an initially maximally entangled state does not always possess the largest negativity after propagating in the GHS dilaton spacetime; instead, under appropriate conditions, certain non-maximally entangled states can retain stronger entanglement than the maximally entangled one. These findings reveal the resource-dependent nature of gravitational effects in dilaton black hole backgrounds and provide new insights into the manipulation and protection of quantum resources for relativistic quantum information processing in curved spacetime.
\end{abstract}

\vspace*{0.5cm}
 \pacs{04.70.Dy, 03.65.Ud,04.62.+v }
\maketitle
\section{Introduction}
Quantum coherence, arising from the superposition principle of quantum mechanics, is one of the most fundamental signatures of quantum behavior \cite{L1,L2}. As a basis-dependent property, it characterizes the ability of a quantum state to preserve coherent superpositions with respect to a chosen reference basis \cite{L47}. Owing to this intrinsic feature, quantum coherence has been recognized as an indispensable resource for a wide range of quantum technologies, including quantum computation, quantum metrology, and quantum communication \cite{L4,L5,L6,L7,L8,L9,L10,L11,L12,L13,L13-1}. In contrast, quantum entanglement describes nonclassical correlations shared among spatially separated subsystems and constitutes a key resource for multipartite quantum information processing \cite{N1}. Although both coherence and entanglement reflect the nonclassical nature of quantum systems, they embody fundamentally different physical concepts: coherence is an intrinsic, basis-dependent property of an individual quantum system \cite{L2}, whereas entanglement characterizes correlations between distinct subsystems and is invariant under local unitary transformations \cite{N1}. Nevertheless, these two quantum resources are intimately connected. In particular, nonlocal coherence can be transformed into quantum entanglement, while coherence distributed across multiple subsystems can itself be regarded as a manifestation of nonclassical correlations. From this perspective, coherence can be viewed as a more fundamental quantum resource, with entanglement emerging as one of its multipartite manifestations \cite{L14,W1}. Despite considerable advances in understanding the interplay between coherence and entanglement, their distinct dynamical characteristics in different physical scenarios, together with their deeper structural relationship, remain incompletely understood. Addressing these open questions is essential not only for establishing a unified framework of quantum resources but also for advancing the theoretical foundations of quantum information science.

Quantum information in curved spacetime has become an important research direction at the intersection of quantum information theory, quantum field theory, and general relativity, offering a natural framework for exploring the behavior of quantum resources under gravitational effects \cite{SDF1,SDF2,SDF3,SDF4,SDF5,SDF6,SDF7,SDF8,SDF9,SDF10,SDF11,SDF12,SDF13,SDF14,SDF15,SDF16,SDF17,SDF18,SDF19,SDF20,SDF21,SDF22,SDF23,SDF24,SDF25,SDF27,SDF28,SDF29,SDF30,SDF31,SDF32,SDF33,SDF34,SDF35,SDF36,SDF37,SDF38,SDF39,SDF40,SDF41,SDF42,SDF43,SDF44,SDF45,SDF46,SDF47,SDF48,SDF49,SDF50,SDF51,SDF52,SDF53,SDF54,
SDF55,SDF56,AGL1,AGL2,AAA1,QFM1,QFM2,QFM3,QFM4,QFM5,ZZZ}. Among various gravitational backgrounds, black hole spacetimes provide ideal theoretical laboratories for investigating the interplay between spacetime curvature and quantum information. In particular, the dilaton black hole, which originates from the low-energy effective theory of string theory, is characterized by a nontrivial dilaton field that modifies the spacetime geometry and consequently influences the evolution of quantum fields  \cite{J9,J10,J11}. As a result, quantum systems propagating in this background can exhibit behaviors that differ significantly from those in flat spacetime \cite{WWW1,WWW2,WWW3,WWW4,WWW5,WWW6,WWW7,WWW8,WWW9,WWW10}. Generally, quantum entanglement and quantum coherence represent fundamentally different physical resources: the former characterizes nonlocal correlations shared among different subsystems, whereas the latter quantifies the superposition of quantum states with respect to a chosen reference basis. Given their distinct physical origins, it remains unclear whether they necessarily respond in the same manner to gravitational effects. This naturally leads to an intriguing question: can the gravitational field of the GHS dilaton black hole, together with an appropriate choice of the initial-state parameters, induce opposite behaviors of different quantum resources, such as suppressing quantum entanglement while simultaneously enhancing quantum coherence? Addressing this issue would not only deepen our understanding of the distinct roles of different quantum resources in curved spacetime but also provide new insights into the development of a unified framework for quantum information in gravitational backgrounds.

To explore the influence of dilaton gravity on different quantum resources, we study a massless Dirac field in the background of the GHS dilaton black hole beyond the single-mode approximation. Quantum entanglement is quantified by the negativity, whereas quantum coherence is characterized by the $l_1$-norm and the relative entropy of coherence (REC). We consider an initially entangled bipartite state shared by Alice and Bob, in which Alice remains in the asymptotically flat region while Bob is located outside the event horizon. By tracing over the field modes inside the horizon, which are inaccessible to external observers, we obtain the reduced density matrix describing the observable subsystem. We demonstrate that the dilaton field gives rise to markedly different responses in quantum entanglement and quantum coherence. Specifically, increasing the dilaton parameter monotonically suppresses entanglement while continuously enhancing coherence, indicating that dilaton-induced spacetime influences nonlocal quantum correlations and local quantum superposition in fundamentally different ways. Furthermore, our results reveal that the response of quantum resources to gravity is strongly
resource-dependent, providing a deeper understanding of quantum information in string-inspired black hole spacetimes and offering useful guidance for protecting and exploiting quantum resources in relativistic quantum information processing.

The remainder of this paper is organized as follows. In Sec. II, we briefly review the quantization of the Dirac field in the background of the GHS dilaton black hole beyond the single-mode approximation and introduce the corresponding vacuum structure. In Sec. III, we systematically investigate the effects of the dilaton black hole on quantum entanglement and quantum coherence, quantifying entanglement by the negativity and coherence by the $l_{1}$-norm and the REC, and present the corresponding numerical results together with their physical interpretations. Finally, Sec. IV summarizes the main conclusions.

\section{Quantization of the Dirac field in a dilaton black hole background}
Low-energy effective actions arising from string theory provide a natural framework for incorporating scalar degrees of freedom coupled to gravity. Among the associated scalar fields, the dilaton field plays a central role and typically couples exponentially to curvature invariants or gauge-field sectors. When the dilaton is nonminimally coupled to the electromagnetic field, one obtains a class of charged black hole solutions that differ qualitatively from their counterparts in Einstein-Maxwell theory. A particularly well-known example is the GHS dilaton black hole, which emerges in the low-energy limit of heterotic string theory. In the Einstein frame, the spacetime geometry of a static, spherically symmetric GHS dilaton black hole is described by the line element
\begin{eqnarray}\label{Q1}
ds^2=-\left(\frac{r-2M}{r-2D}\right)dt^2+\left(\frac{r-2M}{r-2D}\right)^{-1}
 dr^2+r(r-2D)d
 \Omega^2.
\end{eqnarray}
Here $M$ denotes the mass of the black hole, while $D$ characterizes the strength of the dilaton field. These parameters are not independent but are related to the electric charge $Q$ through $D=Q^2/{2M}$. Throughout this work, we adopt natural units by setting $\hbar = G = c = k_{\mathrm{B}} = 1$, unless otherwise stated. The Dirac equation is given by
\begin{eqnarray}\label{Q3}
&&-\frac{\gamma_0}{\sqrt{f}}\frac{\partial \Phi}{\partial t}+\gamma_1\sqrt{f}\bigg[\frac{\partial}{\partial r}+\frac{r-D}{r \bar r}+\frac{1}{4f}\frac{df}{dr} \bigg]\Phi \nonumber\\
&&+\frac{\gamma_2}{\sqrt{r\bar r}}(\frac{\partial}{\partial \theta}+\frac{\cot \theta}{2})\Phi+\frac{\gamma_3}{\sqrt{r\bar r}\sin\theta}\frac{\partial\Phi}{\partial\varphi}=0,
\end{eqnarray}
where $f=\frac{r-2M}{\bar r}$ with $\bar r=r-2D$ \cite{WWW6,WWW8}.  Solving this equation near the event horizon yields the positive-frequency outgoing modes in the exterior and interior regions
\begin{eqnarray}\label{Q4}
\Phi^+_{{\bold k},{\rm in}}\sim \mathcal{R} e^{i\omega \mathcal{H}},
\end{eqnarray}
\begin{eqnarray}\label{Q5}
\Phi^+_{{\bold k},{\rm out}}\sim \mathcal{R} e^{-i\omega \mathcal{H}},
\end{eqnarray}
where $\mathcal{R}$  denotes a four-component Dirac spinor,  $\omega$ is the mode frequency, and $\mathcal{H}=t-r_{*}$ is defined in terms of the tortoise coordinate $r_{*}$. The label
$\boldsymbol k$ collectively denotes the quantum numbers characterizing each mode. Using Eqs. (\ref{Q4}) and (\ref{Q5}), the Dirac field $\Phi$ can be expanded as
\begin{eqnarray}\label{5}
\Phi &=& \int \text{d} \bm{\mathit{k}} [\hat{a}^{\rm out}_{\bm{\mathit{k}}}\Phi^{+}_{\rm out, \bm{\mathit{k}}}+\hat{b}^{\rm out\dag}_{\bm{-\mathit{k}}}\Phi^{-}_{\rm out, \bm{\mathit{k}}}+\hat{a}^{\rm in}_{\bm{\mathit{k}}}\Phi^{+}_{\rm in, \bm{\mathit{k}}}+\hat{b}^{\rm in\dag}_{\bm{-\mathit{k}}}\Phi^{-}_{\rm in, \bm{\mathit{k}}}],
\end{eqnarray}
where $a^{\text{out}}_{\bm{k}}$ and $b^{\text{out}\,\dagger}_{\bm{k}}$ denote the fermionic annihilation and antifermionic creation operators associated with exterior modes, and $a^{\text{in}}_{\bm{k}}$ and $b^{\text{in}\,\dagger}_{\bm{k}}$ are their counterparts for interior modes.

Following the method developed by Damour and Ruffini \cite{WWW11}, the Kruskal modes can be obtained by analytically extending Eqs. (\ref{Q4}) and (\ref{Q5}) across the event horizon. Nevertheless, a direct correspondence between single-frequency Kruskal modes and the fermionic field modes defined in the GHS dilaton spacetime does not generally exist, since a Kruskal observer is, in principle, sensitive to the complete set of accessible field modes rather than to a single-frequency mode defined in the GHS background. To establish the connection between these two descriptions, we introduce the Unruh modes \cite{WWW12}, which serve as an intermediate basis linking the Kruskal and GHS representations. Within this framework, the Unruh creation and annihilation operators are related to the fermionic operators associated with the exterior and interior regions of the GHS dilaton black hole through the following Bogoliubov transformations
\begin{eqnarray}\label{6}
\tilde{c}_{\bm{\mathit{k}},R}&=&\frac{1}{\sqrt{e^{-8\pi (M-D)\omega}+1}}\hat{a}^{\rm out}_{\bm{\mathit{k}}}-\frac{1}{\sqrt{e^{8\pi (M-D)\omega}+1}}\hat{b}^{\rm in\dag}_{\bm{\mathit{-k}}},\nonumber\\
\tilde{c}_{\bm{\mathit{k}},L}&=&\frac{1}{\sqrt{e^{-8\pi (M-D)\omega}+1}}\hat{a}^{\rm in}_{\bm{\mathit{k}}}-\frac{1}{\sqrt{e^{8\pi (M-D)\omega}+1}}\hat{b}^{\rm out\dag}_{\bm{\mathit{-k}}},\nonumber\\
\tilde{c}^{\dag}_{\bm{\mathit{k}},R}&=&\frac{1}{\sqrt{e^{-8\pi (M-D)\omega}+1}}\hat{a}^{\rm out\dag}_{\bm{\mathit{k}}}-\frac{1}{\sqrt{e^{8\pi (M-D)\omega}+1}}\hat{b}^{\rm in}_{\bm{\mathit{-k}}},\nonumber\\
\tilde{c}^{\dag}_{\bm{\mathit{k}},L}&=&\frac{1}{\sqrt{e^{-8\pi (M-D)\omega}+1}}\hat{a}^{\rm in\dag}_{\bm{\mathit{k}}}-\frac{1}{\sqrt{e^{8\pi (M-D)\omega}+1}}\hat{b}^{\rm out}_{\bm{\mathit{-k}}}.
\end{eqnarray}

Here, the subscripts $R$ and $L$ denote the ``right'' and ``left'' Unruh modes, respectively. Adopting the operator ordering $\hat{a}^{\text{out}}_{\bm{k}}  \hat{b}^{\text{out}}_{\bm{-k}} \hat{a}^{\text{in}}_{\bm{k}} \hat{b}^{\text{in}}_{\bm{-k}} $, the Unruh vacuum takes the form
\begin{eqnarray}\label{7}
|0\rangle_{\rm U}&=&\frac{1}{e^{-8\pi (M-D)\omega}+1}|0000\rangle-\frac{1}{\sqrt{e^{8\pi (M-D)\omega}+e^{-8\pi (M-D)\omega}+2}}|0110\rangle\nonumber\\
&+&\frac{1}{\sqrt{e^{8\pi (M-D)\omega}+e^{-8\pi (M-D)\omega}+2}}|1001\rangle-\frac{1}{e^{8\pi (M-D)\omega}+1}|1111\rangle,
\end{eqnarray}
and
\begin{eqnarray}\label{8}
|mnm'n'\rangle=|m_{\bm{\mathit{k}}}\rangle^{+}_{\rm out}|n_{\bm{\mathit{-k}}}\rangle^{-}_{\rm out}|m'_{\bm{\mathit{k}}}\rangle^{+}_{\rm in}|n'_{\bm{\mathit{-k}}}\rangle^{-}_{\rm in}.
\end{eqnarray}
Here, the basis states ${|n_{\bm{\mathit{k}}}\rangle^{+}_{\mathrm{out}}}$ and ${|n_{-\bm{\mathit{k}}}\rangle^{-}_{\mathrm{in}}}$ constitute complete orthonormal bases for the exterior and interior regions of the GHS dilaton black hole, respectively, where the superscripts $\{+,-\}$ denote the fermionic and antifermionic sectors. From the perspective of an observer outside the event horizon, the expected particle number follows a Fermi-Dirac distribution
\[
N_F=\frac{1}{e^{8\pi(M-D)\omega}+1},
\]
where the dilaton parameter \rev{$D$} modifies the particle distribution through the geometry of the GHS spacetime. Furthermore, because of the Pauli exclusion principle, each fermionic mode can be occupied by at most one particle. Accordingly, the Unruh single-particle state takes the form
\begin{eqnarray}\label{10}
|1\rangle_{\rm U}&=&[q_{R}(\tilde{c}^{\dag}_{\bm{k},R}\bigotimes I_{L})+q_{L}(I_{R}\bigotimes\tilde{c}^{\dag}_{\bm{k},L})]|0\rangle_{\rm U}\nonumber\\
&=&q_{R}[\frac{1}{\sqrt{e^{-8\pi (M-D)\omega+1}}}|1000\rangle-\frac{1}{\sqrt{e^{8\pi (M-D)\omega+1}}}|1110\rangle]\nonumber\\
&&+q_{L}[\frac{1}{\sqrt{e^{-8\pi (M-D)\omega+1}}}|0010\rangle+\frac{1}{\sqrt{e^{8\pi (M-D)\omega+1}}}|1011\rangle],
\end{eqnarray}
with the normalization condition $|q_{R}|^2+|q_{L}|^2=1$.
From a theoretical viewpoint, this normalization condition guarantees that the Bogoliubov transformation relating the GHS dilaton black hole modes to the Unruh modes preserves the canonical fermionic anticommutation relations. Accordingly, the coefficients $q_R$ and $q_L$ constitute a normalized pair of amplitudes satisfying $|q_R|^2+|q_L|^2=1.$ These parameters characterize the probability amplitudes associated with excitations in the right and left Unruh sectors, respectively. Since the two sectors represent mutually exclusive excitation channels, the corresponding one-particle state must remain properly normalized, giving rise to the above constraint. Physically, this relation indicates that a detector interacting with the field can excite either the right or the left Unruh mode. The limiting case $q_R=1$ (equivalently $q_L=0$) recovers the conventional single-mode approximation, while arbitrary normalized values of $q_R$ and $q_L$ describe the more general scenario beyond the single-mode approximation.

\section{Influence of the dilaton-induced gravitational field on quantum entanglement and quantum coherence}
We consider a bipartite system in which two observers, Alice and Bob, initially share an entangled fermionic state. The initial pure state can be written as
\begin{equation}
|\Phi\rangle=\cos\beta\,|0\rangle_{M}|0\rangle_{U}+\sin\beta\,|1\rangle_{M}|1\rangle_{U},
\end{equation}
where the subscripts $M$ and $U$ label the Minkowski and Unruh modes accessible in the asymptotic region, respectively.

\begin{figure}[htbp]
\centering
\includegraphics[height=2in,width=4in]{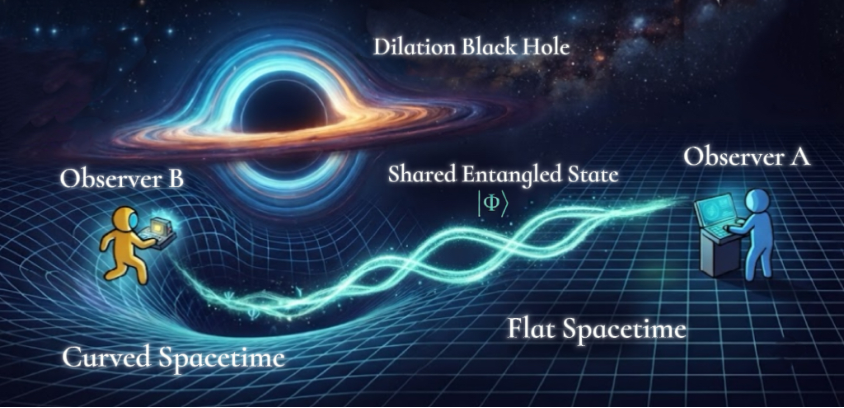}
\caption{Schematic illustration of the physical setup in the GHS dilaton black hole spacetime. Alice is located in the asymptotically flat region, whereas Bob remains static outside the event horizon. Owing to the causal separation imposed by the event horizon, Bob has access only to the quantum field modes in the exterior region.}
\label{F1}
\end{figure}

As illustrated in Fig. \ref{F1}, Alice remains in the asymptotically flat region, where the spacetime is effectively Minkowskian and her detector is naturally described by Minkowski modes. Bob, on the other hand, stays at a fixed radial position outside the event horizon of the GHS dilaton black hole and performs local measurements on the quantum field in the exterior region. The event horizon divides the spacetime into causally disconnected interior and exterior regions, implying that an observer outside the horizon can access only the field degrees of freedom in the exterior region. Accordingly, to describe the quantum state shared by Alice and Bob, the field is decomposed into modes associated with the exterior and interior regions of the event horizon. Since the interior modes are inaccessible to Bob, they are regarded as unobservable degrees of freedom and are traced out when constructing the reduced density operator accessible to the two observers. The reduced density matrix accessible to Alice and Bob is obtained as
\begin{align}\label{13}
\rho^{\Phi}_{AB} &=
\cos^{2}\beta \left( \dfrac{1}{1+\text{e}^{-8\pi (M-D)\omega}} \right)^{2} |000\rangle\langle 000| \notag\\
&\quad + \dfrac{q_R}{2} \sin 2\beta
\left( \dfrac{1}{1+\text{e}^{-8\pi (M-D)\omega}} \right)^{\frac{3}{2}}
(|000\rangle\langle 110| + |110\rangle\langle 000|) \notag\\
&\quad + q_L^{2} \sin^{2}\beta \dfrac{1}{1+\text{e}^{-8\pi (M-D)\omega}} |100\rangle\langle 100| \notag\\
&\quad + \dfrac{1}{2}\Big[1+(1-2q_L^{2})(\dfrac{2}{1+\text{e}^{-8\pi (M-D)\omega}}-1)\Big]\sin^{2}\beta |110\rangle\langle 110| \notag\\
&\quad - \dfrac{q_L}{2} \sin 2\beta \dfrac{1}{1+\text{e}^{-8\pi (M-D)\omega}}
\dfrac{1}{\sqrt{1+\text{e}^{8\pi (M-D)\omega}}} (|001\rangle\langle 100| + |100\rangle\langle 001|) \notag\\
&\quad - q_R q_L\sin^{2}\beta \dfrac{1}{\sqrt{(1+\text{e}^{-8\pi (M-D)\omega})(1+\text{e}^{8\pi (M-D)\omega})}} (|100\rangle\langle 111| + |111\rangle\langle 100|) \notag\\
&\quad + \cos^{2}\beta \dfrac{1}{1+\text{e}^{-8\pi (M-D)\omega}}\dfrac{1}{1+\text{e}^{8\pi (M-D)\omega}} (|001\rangle\langle 001| + |010\rangle\langle 010|) \notag\\
&\quad + \dfrac{q_R}{2}\sin 2\beta
\dfrac{1}{\sqrt{1+\text{e}^{-8\pi (M-D)\omega}}}
\dfrac{1}{1+\text{e}^{8\pi (M-D)\omega}} (|001\rangle\langle 111| + |111\rangle\langle 001|) \notag\\
&\quad + q_R^{2}\sin^{2}\beta \dfrac{1}{1+\text{e}^{8\pi (M-D)\omega}} |111\rangle\langle 111| \notag\\
&\quad - \dfrac{q_L}{2}\sin 2\beta
\left(\dfrac{1}{\sqrt{1+\text{e}^{8\pi (M-D)\omega}}}\right)^{\frac{3}{2}}
(|011\rangle\langle 110| + |110\rangle\langle 011|) \notag\\
&\quad + \cos^{2}\beta \left( \dfrac{1}{1+\text{e}^{8\pi (M-D)\omega}} \right)^{2} |011\rangle\langle 011|,
\end{align}
where the basis used is $|ijk\rangle = \overbrace{|i\rangle_{M}}^{\text{Alice}} \overbrace{|j\rangle_{\text {out}}^+ |k\rangle_{\text {out}}^-}^{\text{Bob}}.$

To simplify the notation, we introduce the quantities $\cos^2\gamma = \frac{1}{1 + \text{e}^{-8\pi (M-D)\omega}}, \quad\sin^2\gamma = \frac{1}{1 + \text{e}^{8\pi (M-D)\omega}}, \quad\sin\gamma\cos\gamma = \frac{1}{\sqrt{(1 + \text{e}^{-8\pi (M-D)\omega})(1 + \text{e}^{8\pi (M-D)\omega})}}$,
which considerably simplify the algebraic structure of the reduced density operator. Using this notation, the reduced density operator can be written in matrix form as
\[
\scalebox{0.5}{$
\rho^{\Phi}_{AB} =
\begin{pmatrix}
\cos^2\beta \cos^4\gamma & 0 & 0 & 0 & 0 & 0 & \tfrac{q_R}{2}\sin 2\beta \cos^3\gamma & 0 \\[6pt]
0 & \frac{1}{4}\cos^2\beta \sin^2 2\gamma & 0 & 0 & -\tfrac{q_L}{2}\sin 2\beta \sin\gamma \cos^2\gamma & 0 & 0 & \tfrac{q_R}{2}\sin 2\beta \sin^2\gamma \cos\gamma \\[6pt]
0 & 0 & \frac{1}{4}\cos^2\beta \sin^2 2\gamma & 0 & 0 & 0 & 0 & 0 \\[6pt]
0 & 0 & 0 & \cos^2\beta \sin^4\gamma & 0 & 0 & -\tfrac{q_L}{2}\sin 2\beta \sin^3\gamma & 0 \\[6pt]
0 & -\tfrac{q_L}{2}\sin 2\beta \sin\gamma \cos^2\gamma & 0 & 0 & q_L^2 \sin^2\beta \cos^2\gamma & 0 & 0 & -q_R q_L \sin^2\beta \sin\gamma \cos\gamma \\[6pt]
0 & 0 & 0 & 0 & 0 & 0 & 0 & 0 \\[6pt]
\tfrac{q_R}{2}\sin 2\beta \cos^3\gamma  & 0 & 0 & -\tfrac{q_L}{2}\sin 2\beta \sin^3\gamma & 0 & 0 & \tfrac{1}{2}\big[1+(1-2q_L^2)(2\cos^2\gamma-1)\big]\sin^2\beta & 0 \\[6pt]
0 & \tfrac{q_R}{2}\sin 2\beta \sin^2\gamma \cos\gamma & 0 & 0 & -q_R q_L \sin^2\beta \sin\gamma \cos\gamma & 0 & 0 & q_R^2 \sin^2\beta \sin^2\gamma .
\end{pmatrix}
$}
\]

In this work, we characterize the quantum coherence of the detector state by employing two widely used quantifiers, namely the $l_1$-norm of coherence and the REC \cite{L47}. Since quantum coherence is inherently basis dependent, we adopt the occupation-number basis introduced above as the natural reference basis, which corresponds to particle excitations accessible to observers outside the GHS dilaton black hole. More generally, for an $n$-dimensional quantum system described by a density matrix $\rho$ in the reference basis $\{|i\rangle\}_{i=1,\ldots,n}$, the $l_1$-norm of coherence is defined as the sum of the absolute values of all off-diagonal elements of $\rho$ \cite{L47}, namely
\begin{eqnarray}\label{14}
C_{l_{1}}=\sum_{i\neq j}|\rho_{i,j}|.
\end{eqnarray}
Substituting the reduced density matrix \( \rho^{\Phi}_{AB} \) into Eq. (\ref{14}), the \( l_1 \)-norm of quantum coherence can be evaluated explicitly and is given by
\begin{equation}
\begin{aligned}\label{16}
C_{l_1}(\rho^{\Phi}_{AB})&=q_L q_R \frac{2}{\sqrt{(1+\text{e}^{-8\pi (M-D)\omega})(1+\text{e}^{8\pi (M-D)\omega})}} \sin^2\beta + q_L \frac{1}{\sqrt{1+\text{e}^{8\pi (M-D)\omega}}} \sin2\beta \\
&+ q_R \frac{1}{1+\text{e}^{8\pi (M-D)\omega}} \sqrt{\frac{1}{1+\text{e}^{-8\pi (M-D)\omega}}} \sin2\beta + q_R \left( \frac{1}{1+\text{e}^{-8\pi (M-D)\omega}} \right)^{3/2} \sin2\beta.
\end{aligned}
\end{equation}
In addition to the $l_{1}$-norm of coherence, we employ the REC, which quantifies quantum coherence through the entropy difference between a quantum state and its corresponding incoherent state. For the reduced density matrix $\rho_{AB}^{\Phi}$, the REC is defined as
\begin{eqnarray}\label{15}
C_{\textrm{RE}}(\rho^{\Phi}_{AB}) = S(\rho^{\Phi}_{AB,\textrm{diag}}) -S(\rho^{\Phi}_{AB}),
\end{eqnarray}
where $\rho^{\Phi}_{AB,\mathrm{diag}}$ denotes the diagonal part of $\rho^{\Phi}_{AB}$ in the chosen reference basis, and $S(\rho^{\Phi}_{AB})$ is the von Neumann entropy of $\rho^{\Phi}_{AB}$. Owing to the block-diagonal structure of the reduced density matrix, only five eigenvalues are nonzero. Since their analytical expressions are rather lengthy and offer little additional physical insight, we do not present them explicitly.

The entanglement of the mixed bipartite state is quantified by the negativity, which provides an effective measure of quantum entanglement through the spectrum of the partially transposed density matrix \cite{WWW13}. For a bipartite state $\rho_{AB}$, the negativity is defined as
\begin{eqnarray}\label{18}
N(\rho_{AB})  = \left\| \rho_{AB}^{T_A} \right\| - 1.
\end{eqnarray}
Here, $T_A$ denotes the partial transpose with respect to subsystem $A$. Since $\left\| \rho_{AB}^{T_A} \right\| - 1$ equals twice the sum of the absolute values of the negative eigenvalues, the entanglement can alternatively be expressed as
\begin{eqnarray} \label{19}
N(\rho^{\Phi}_{AB}) = 2 \sum_{i=1}^{n} \left| \lambda^{-}_{ \left( \rho^{\Phi,T_{A}}_{AB} \right)} \right|^{i},
\end{eqnarray}
where $\lambda^{-}_{\left(\rho^{\Phi,T_{A}}_{AB}\right)}$ denote the negative eigenvalues of the partially transposed matrix.

\begin{figure}[t]
\centering
\includegraphics[width=0.85\linewidth]{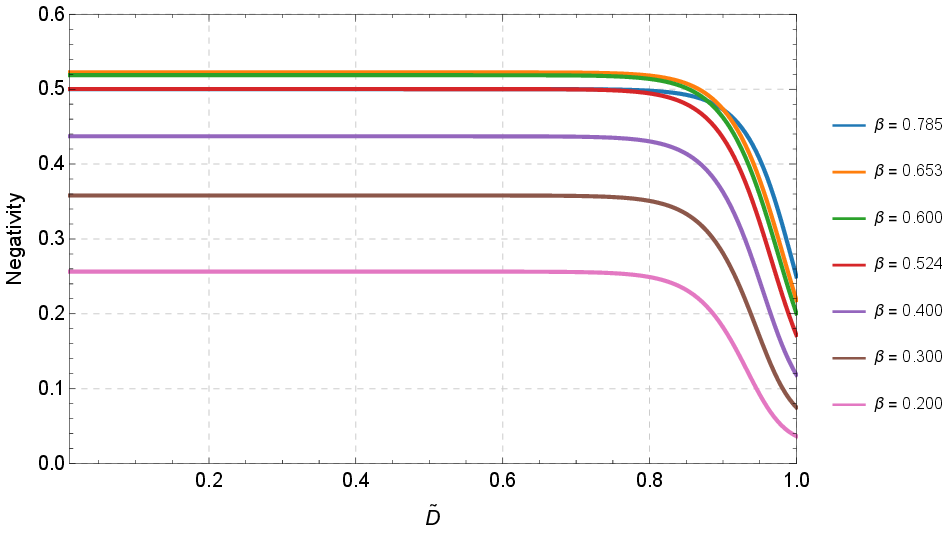}
\caption{$N(\rho_{AB}^{\Phi})$ as a function of the dilaton parameter $D$ for several representative values of the initial-state parameter $\beta$. The remaining parameters are fixed at $\omega=1$ and $q_{R}=1/\sqrt{2}$. The colored curves correspond to $\beta=0.785$ (blue), $0.653$ (orange), $0.600$ (green), $0.524$ (red), $0.400$ (purple), $0.300$ (brown), and $0.200$ (pink).}
\label{F2}
\end{figure}

\begin{figure}[t]
\centering

\includegraphics[width=0.85\linewidth]{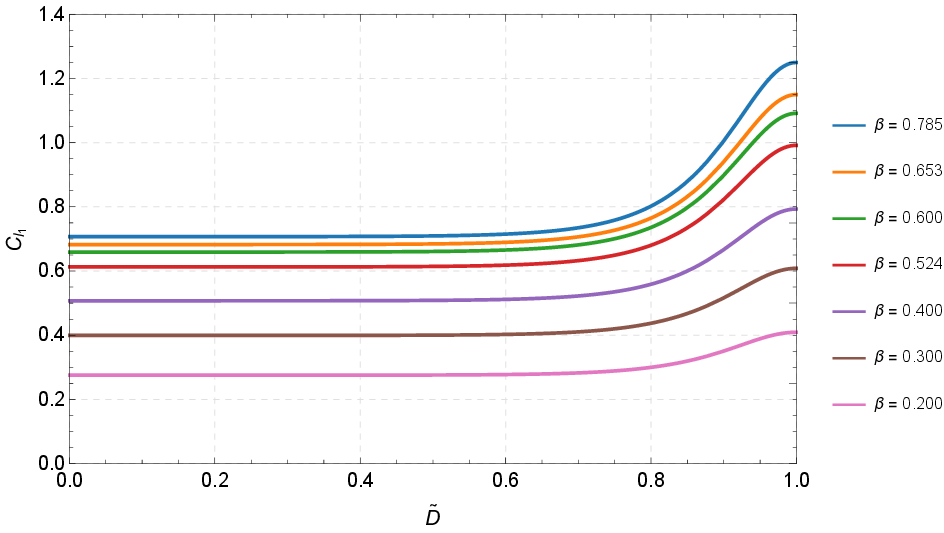}

\vspace{0.4cm}

\includegraphics[width=0.85\linewidth]{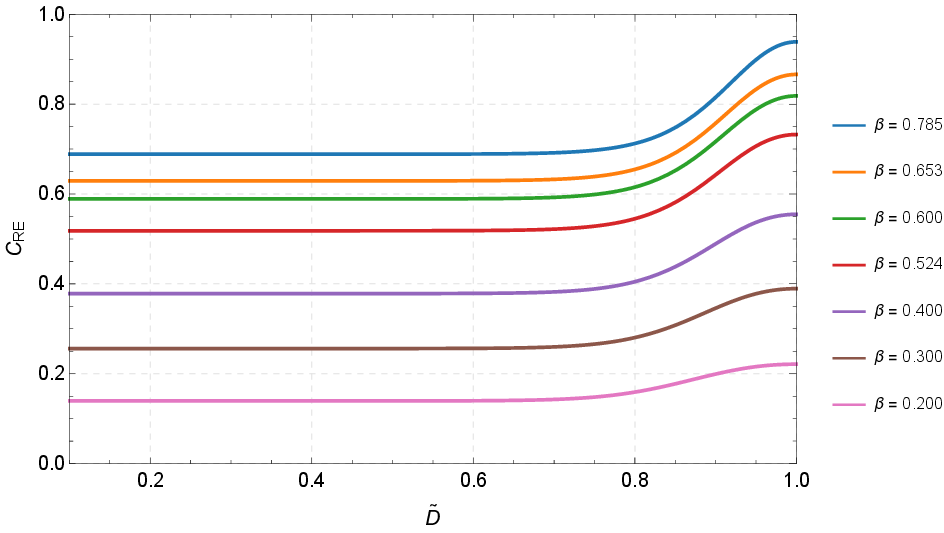}

\caption{Dependence of quantum coherence on the dilaton parameter $D$ for various values of $\beta$. The parameters are fixed at $\omega=1$ and $q_{R}=1/\sqrt{2}$. The colored curves correspond to $\beta=0.785$ (blue), $0.653$ (orange), $0.600$ (green), $0.524$ (red), $0.400$ (purple), $0.300$ (brown), and $0.200$ (pink).}
\label{F3}

\end{figure}

Fig. \ref{F2} illustrates the evolution of the negativity $N(\rho_{AB}^{\Phi})$ as a function of the dilaton parameter $D$ for different initial-state parameters $\beta$. It is evident that the quantum entanglement decreases monotonically with increasing $D$ for all initial states, indicating that the dilaton-induced gravitational field continuously suppresses nonlocal quantum correlations as the gravitational field strengthens. As the dilaton parameter approaches its upper limit, corresponding to the strong-gravity regime, the negativity gradually approaches a finite residual value rather than vanishing completely. This behavior suggests that although the dilaton field significantly weakens quantum entanglement, a portion of the entanglement remains robust against gravitational effects. Another noteworthy feature is that the maximally entangled initial state $\beta=\frac{\pi}{4}$ does not always exhibit the largest observable negativity. Instead, for sufficiently small values of the dilaton parameter, certain non-maximally entangled states possess a larger negativity than the maximally entangled one. This demonstrates that the observable entanglement in the GHS dilaton spacetime is determined not only by the amount of initial entanglement but also by how the initial state responds to the dilaton-induced modification of the spacetime geometry. As the gravitational field becomes stronger, however, the maximally entangled state gradually becomes the most robust one and eventually retains the largest residual entanglement in the strong-gravity regime. These results indicate that the survivability of quantum entanglement in curved spacetime depends sensitively on both the initial-state structure and the strength of the dilaton field.

In sharp contrast to the behavior of quantum entanglement shown in Fig. \ref{F2}, both the $l_{1}$-norm of coherence and the REC increase monotonically with the dilaton parameter, as illustrated in Fig. \ref{F3}. This opposite trend reveals that the dilaton-induced gravitational field affects different quantum resources in fundamentally different ways. While quantum entanglement, which characterizes nonlocal correlations between subsystems, is continuously suppressed by the increasing gravitational influence, quantum coherence, which describes local quantum superposition with respect to the chosen reference basis, is instead enhanced throughout the evolution. Moreover, the ordering of the coherence curves remains unchanged during the entire evolution. States with larger initial coherence always maintain larger coherence for every value of the dilaton parameter, demonstrating that quantum coherence is remarkably robust against the influence of the dilaton field. Unlike quantum entanglement, whose robustness strongly depends on the detailed structure of the initial state, coherence exhibits a stable and predictable evolution under the same gravitational background. The opposite monotonic behaviors of entanglement and coherence clearly demonstrate that the influence of the GHS dilaton black hole is strongly resource-dependent. Although both quantities originate from quantum superposition, they characterize different aspects of quantumness and therefore respond differently to the modification of spacetime induced by the dilaton field. These results further indicate that protecting quantum entanglement and exploiting quantum coherence require different strategies in relativistic quantum information processing performed in dilaton black-hole spacetimes.

\section{Conclusions}
In this work, we have investigated the influence of the GHS dilaton black hole on the quantum entanglement and coherence of Dirac fields beyond the single-mode approximation. Quantum entanglement was quantified by the negativity, whereas coherence was quantified by the $l_{1}$-norm of coherence and the REC. By tracing over the inaccessible field modes inside the event horizon, we analyzed the evolution of different quantum resources in the presence of the dilaton-induced gravitational field. Our results reveal a clear distinction between the behaviors of entanglement and coherence. As the dilaton parameter increases, quantum entanglement is monotonically suppressed and eventually approaches a finite residual value, whereas both coherence measures are continuously enhanced. This demonstrates that the dilaton field affects nonlocal quantum correlations and local quantum superposition in fundamentally different ways. Furthermore, we find that the amount of initial entanglement alone does not determine the robustness of a quantum state in curved spacetime. The evolution of quantum resources depends sensitively on the structure of the initial state, indicating that states with different initial configurations can exhibit markedly different survivability under gravitational effects. In contrast, quantum coherence remains remarkably robust and can even be amplified by the dilaton-induced spacetime. Overall, our findings highlight the resource-dependent nature of gravitational effects and provide further insights into the behavior of quantum information in string-inspired black hole backgrounds. These results may be useful for the development of relativistic quantum information protocols and for understanding the interplay between gravity and different manifestations of quantumness.

\begin{acknowledgments}
This work was supported by the Fundamental Research Program of Shanxi Province under Grant No.202503021212266, the Scientific and Technological Innovation Programs of Higher Education Institutions of Shanxi Province under Grant No. 2025L120, the Liaoning Provincial Doctoral Scientific Research Startup Foundation under Grant No.2024-BS-286 and the National Natural Science Foundation of China (Grant No. 12574384).	
\end{acknowledgments}

\end{document}